\documentclass[fleqn,usenatbib,useAMS]{mnras}

\usepackage{newtxtext,newtxmath}
\usepackage{subfig}
\usepackage{upgreek}
\usepackage{soul}

\usepackage[T1]{fontenc}
\newcommand{\Msun}{\,${\rm M}_\odot$\,}
\DeclareRobustCommand{\VAN}[3]{#2}
\let\VANthebibliography\thebibliography
\def\thebibliography{\DeclareRobustCommand{\VAN}[3]{##3}\VANthebibliography}

\usepackage{graphicx}	
\usepackage{amsmath}	

\def\gsim{ \lower .75ex \hbox{$\sim$} \llap{\raise .27ex \hbox{$>$}} }
\def\lsim{ \lower .75ex \hbox{$\sim$} \llap{\raise .27ex \hbox{$<$}} }

\title[{Free-streaming in voids}]{The impact of free-streaming on dwarf galaxy counts in low-density regions}

\author[Tamar Meshveliani]{
Tamar Meshveliani$^{1}$\thanks{E-mail: tam15@hi.is},
Mark R. Lovell$^{1,2,3}$, 
Robert A. Crain$^{4}$ and
Joel Pfeffer$^{5}$
\\
$^{1}$Centre for Astrophysics and Cosmology, Science Institute,
University of Iceland, Dunhagi 5, 107 Reykjavik, Iceland \\
$^{2}$ Institute for Computational Cosmology, Durham University, South Road, Durham DH1 3LE, United Kingdom\\
$^{3}$ Department of Physics, Durham University, South Road, Durham DH1 3LE, United Kingdom \\
$^{4}$Astrophysics Research Institute, Liverpool John Moores University, 146 Brownlow Hill, Liverpool L3 5RF, UK \\
$^{5}$Centre for Astrophysics \& Supercomputing, Swinburne University, Hawthorn, VIC 3122, Australia
}

\date{Accepted XXX. Received YYY; in original form ZZZ}

\pubyear{2015}

\begin{document}
\label{firstpage}
\pagerange{\pageref{firstpage}--\pageref{lastpage}}
\maketitle

\begin{abstract}
We study the statistics of dwarf galaxy populations as a function of environment in the cold dark matter (CDM) and warm dark matter (WDM) cosmogonies, using hydrodynamical simulations starting from initial conditions with matched phases but differing power spectra, and evolved with the EAGLE galaxy formation model. We measure the abundance of dwarf galaxies within 3~Mpc of DM haloes with a present-day halo mass similar to that of the Milky Way (MW), and find that the radial distribution of galaxies $M_{*}>10^7$\Msun is nearly identical for WDM and CDM. However, the cumulative mass function becomes shallower for WDM at lower masses, yielding 50~per~cent fewer dwarf galaxies of $M_{*}\gsim10^{5}$~\Msun than CDM. The suppression of low-mass halo counts in WDM relative to CDM increases significantly from high-density regions to low-density regions for haloes in the region of the half-mode mass, $M_\rmn{hm}$. The luminous fraction in the two models also diverges from the overdense to the underdense regions for $M>2M_\rmn{hm}$, as the increased collapse delay at small densities pushes the collapse to after the reionization threshold. However, the stellar mass--halo mass relation of WDM haloes relative to CDM increases towards lower-density regions. Finally, we conclude that the suppression of galaxies with $M_{*}\gsim10^5$\Msun between WDM and CDM is independent of density: the suppression of halo counts and the luminous fraction is balanced by an enhancement in stellar mass--halo mass relation.

\end{abstract}




\section{Introduction}
\label{sec:intro}
The cold dark matter (CDM) model has been highly successful in explaining the Universe's large-scale structure, including the predictions for the cosmic microwave background \citep{Planck2014} and the large scale distribution of galaxies \citep{Eisenstein2005}. One of its key predictions is the existence of dark matter (DM) particles whose streaming velocities are negligible for most astrophysical considerations. The greatest uncertainty in the zero-streaming velocity paradigm is expected for dwarf galaxies, where the constraints on free-streaming are weakest. It is also in this regime that the role of astrophysics is weakest and so is less degenerate with novel DM physics \citep{DiCintio2014}. Therefore, the dwarf galaxies of the Local Group (LG) and their haloes have become popular test cases for the impact of cosmological models on small-scale structures \citep{Polisensky2011,Lovell2014, Vogelsberger2012, Bozek2016,Horiuchi2016,Cherry2017,Kim2018,Newton2018, Maccio2019,Nadler2020,Enzi2021,Nadler2021}.

For instance, there is a reported mismatch between observations and the predictions of CDM $N$-body simulations for the abundance and kinematic properties of the Milky Way's (MW's) satellites. The inner regions of several bright MW satellite galaxies are measured to be DM-dominated yet are less dense than is predicted by $N$-body CDM simulations of MW-analogue haloes, the well-known `Too Big To Fail' problem \citep{Boylan-kolchin2011, Boylan-Kolchin2012}. Another challenge, known as the cusp-core problem, states that central regions of several MW satellites are more accurately described by roughly constant radial density profiles than the steep inner density slope predicted for standard CDM haloes \citep{Walker2011}. Finally, it is also unclear whether galaxy formation models based on CDM can predict accurately the number of faint, isolated dwarf galaxies in the LG \citep{Kim2018}. It is anticipated that astrophysical processes, including reionization and feedback from the formation and evolution of stars will play a crucial role in shaping the LG dwarf population \citep[e.g.][]{Governato2012,Sawala2016a,Lovell2017a}, but the simulations representing the current state-of-the-art, such as APOSTLE \citep{Fattahi2016} overpredict the number of LG galaxies of stellar mass $>10^{5}$~\Msun by over 50~per~cent \citep{Fattahi2020}.

A further challenge to the CDM model is that none of its particle physics candidates have been directly observed or detected by experiments. Collider searches have yet to show any evidence for DM production \citep{ATLAS2018, CMS2018} and similarly underground direct detection experiments have not identified a conclusive set of DM collision events, either for supersymmetric weakly interacting massive particles \citep{Aprile2018,Lanfranchi2021} or for QCD axions \citep{Rosenberg2015,Du2018}. Taken together with the inconclusive nature of gamma-ray indirect detection efforts \citep{Hooper2011,Albert2017}, the identification of DM in experiments remains an outstanding problem.

Given the dual challenges posed to CDM -- the discrepancies at small scales and the lack of evidence for its DM particle in experiments -- there is strong motivation to consider other DM candidates that can address these challenges simultaneously. One such candidate is the resonantly-produced sterile neutrino \citep{Shi1999}. This candidate is part of a standard model extension called the neutrino Minimal Standard Model \citep[$\nu \rm{MSM}$;][]{Asaka2005,Boyarsky2009}, which is well motivated from a particle physics perspective in that it has the potential to explain baryogenesis and neutrino flavour oscillations as well as supply a DM candidate. It belongs to the warm DM (WDM) subset of DM candidates in that it exhibits a significant primordial velocity distribution, erases small dwarf galaxy-mass perturbations in the early Universe, and subsequently produces a cutoff in the linear fluctuation power spectrum. This cutoff impacts the properties of dwarf galaxies in several ways relevant to the CDM small-scale challenges, including delaying their formation time \citep{Lovell2012}, lowering their central densities \citep{Lovell2012}, and reducing their number density \citep{Colin2000, Bode2001,Polisensky2011}.

Another compelling aspect of the resonantly-produced sterile neutrino is that it has a decay channel that generates an X-ray photon and is, therefore, subject to indirect detection constraints. The decay rate of the sterile neutrinos is set in part by their mixing angle, $\sin^{2}(2\theta)$, which also plays a role in setting the free-streaming length. The detection of an X-ray decay signal, therefore, determines the power spectrum cutoff, enabling this detection to be probed with measurements of dwarf galaxy counts and densities. One such signal is an unexplained 3.55~keV X-ray line reported in galaxy clusters, the M31 galaxy, and the Galactic Centre (\citealp{Bulbul2014, Boyarsky2014, Hofmann2019}; see \citealp{Anderson2015, Jeltema2016, Dessert2020,Dessert2023} for alternative interpretations that include uncertainties in the X-ray background modelling and the proposed contribution of charge exchange). The sterile neutrino decay interpretation is consistent with a mixing angle in the range $[2,20]\times10^{-11}$, which corresponds to a cutoff scale wavenumber $k\sim[7,13]~h\rmn{Mpc}^{-1}$ \citep{Lovell2023}, therefore adopting this model we have a fixed WDM scale at which to test the sterile neutrino as an alternative solution to the CDM small-scale challenges.

Cosmologically underdense regions present an attractive and relatively unexplored regime in which to uncover divergent predictions of CDM and WDM. This regime is of interest because the absence of large scale overdensities means the collapse of DM haloes is governed by small-scale fluctuations, whose absence differentiates WDM from CDM. \citet{Lovell2024} demonstrated explicitly that halo collapse time -- the time at which a halo first becomes massive enough to undergo atomic hydrogen cooling -- is delayed in WDM relative to CDM, and the delay becomes longer for progressively lower-mass haloes. It has been shown that the general delay in WDM halo formation leads to lower halo densities \citep{Lovell2012} and to later reionization times \citep{Bose2016b}; therefore, it may lead to a further suppression over and above that of the halo mass function in voids that could be detected by deep surveys such as Vera C. Rubin Observatory \citep{LSST2019}. This change would be relevant for LG dwarf count analyses \citep[e.g.][]{Fattahi2020}, and we should expect the suppression of dwarf galaxies to differ as a function of the local density between different subregions of the LG.

The relatively recent emergence of galaxy formation models capable of yielding galaxy populations with broadly realistic properties \citep[see, e.g.][and references therein]{Crain2023} affords an exciting opportunity to compare the outcomes of `paired' simulations that use a fixed galaxy formation model, and whose initial conditions have matched phases thus differing only in terms of the adopted power spectrum. Here, we use such simulations to understand the relationship between the statistics and properties of dwarf galaxy populations as a function of the local overdensity in the CDM and WDM cosmogonies, with a particular focus on how this change may alleviate the discrepancy between observed dwarf galaxy counts and the predictions of CDM simulations of the LG highlighted by \citet{Fattahi2020}. We select our WDM model from the set of models favoured by the reported 3.55~keV line, which yields a preferred halo suppression scale independent of structure formation considerations, and conduct a new cosmological hydrodynamical simulation of galaxy formation that is a WDM partner to an existing CDM simulation from the EAGLE project. We consider the impact of changes to the halo mass function, the luminous fraction, and the stellar mass--halo mass relation on the likelihood of detecting faint dwarf galaxies with future surveys. We thus test the hypothesis that the abundance of low-mass galaxies is suppressed even further between CDM and WDM in low-density regions than is the case in high-density regions. This potential disparity could be an additional factor contributing to the absence of dwarf galaxies in the LG (see also \citealp{Bose2016, Lovell2019}). The paper is structured as follows: in Sec.~\ref{sec:methods}, we present the numerical simulations, the LG galaxy sample and describe our methods; in Sec.~\ref{sec:results}, we present our results, and we draw our conclusions in Sec.~\ref{sec:conclusions}.

\section{Methods}
\label{sec:methods}

 In Sec.~\ref{sec:sim}, we provide a brief overview of the simulations we examine. Many aspects of simulations have been described in detail elsewhere, so we restrict ourselves to a brief overview and discuss primarily the novel initial conditions of the WDM simulations. We discuss the WDM model we adopt in   Sec.~\ref{sec:wdm}. We briefly outline our Local Group galaxy sample in Sec.~\ref{sec:sample}. 
 
\begin{figure*}
  \centering
   \setbox1=\hbox{\includegraphics[scale=0.23]{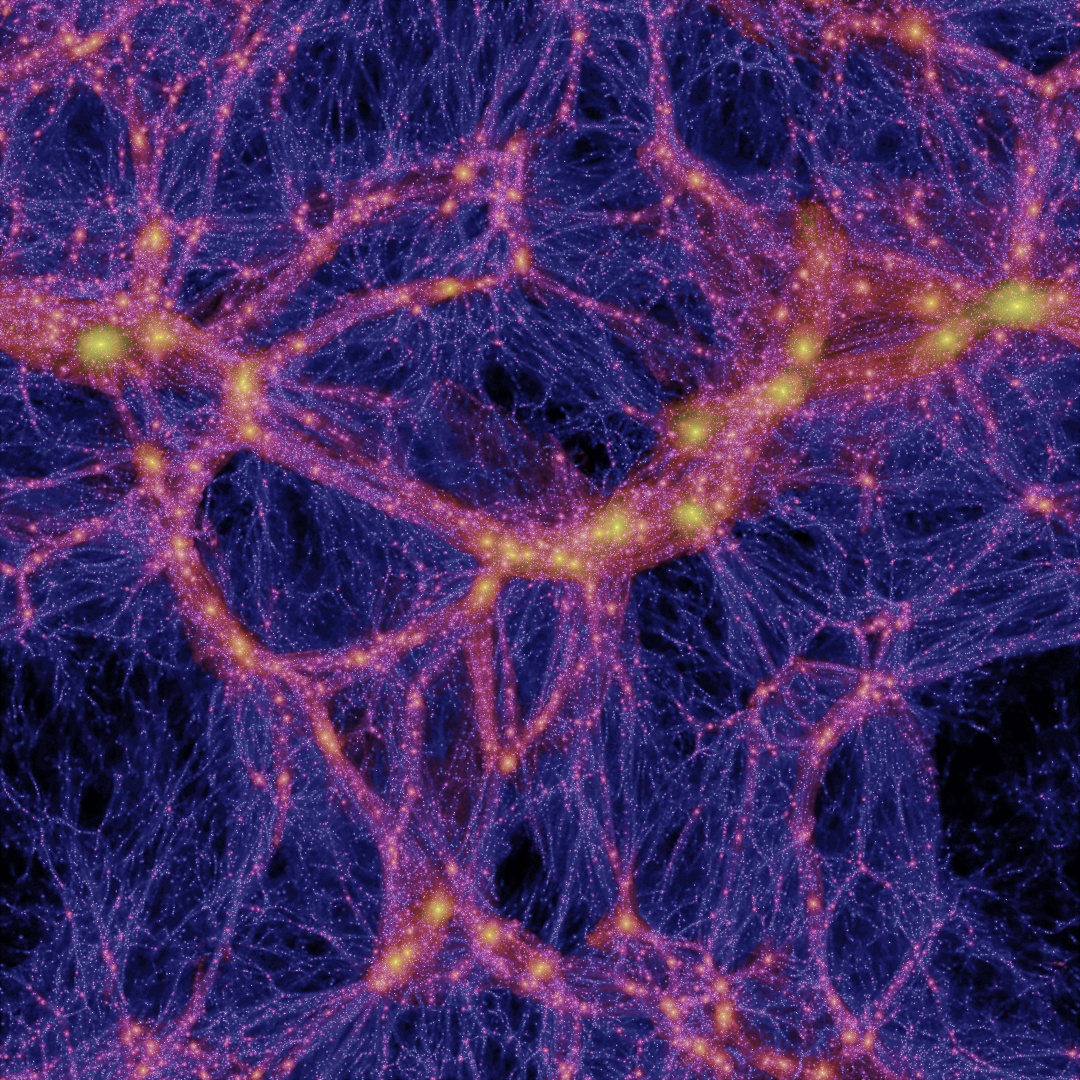}}
  \includegraphics[scale=0.23]{EMOSAICS_L34N1034_28_PType2_Bsize11650kpc_Sthick3000kpc_DensVsVelDisp.jpg}\llap{\makebox[\wd1][l]{\raisebox{-0.1\wd1}{\includegraphics[width=0.36\textwidth]{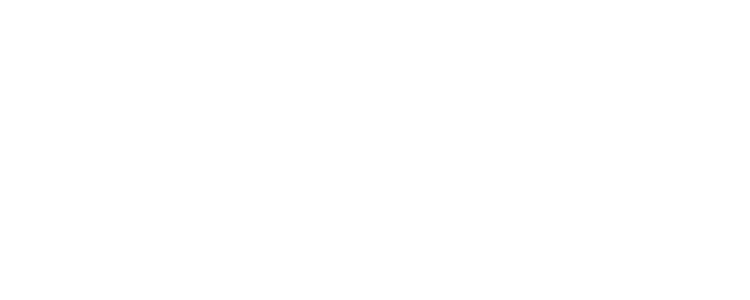}}}}
  \includegraphics[scale=0.23]{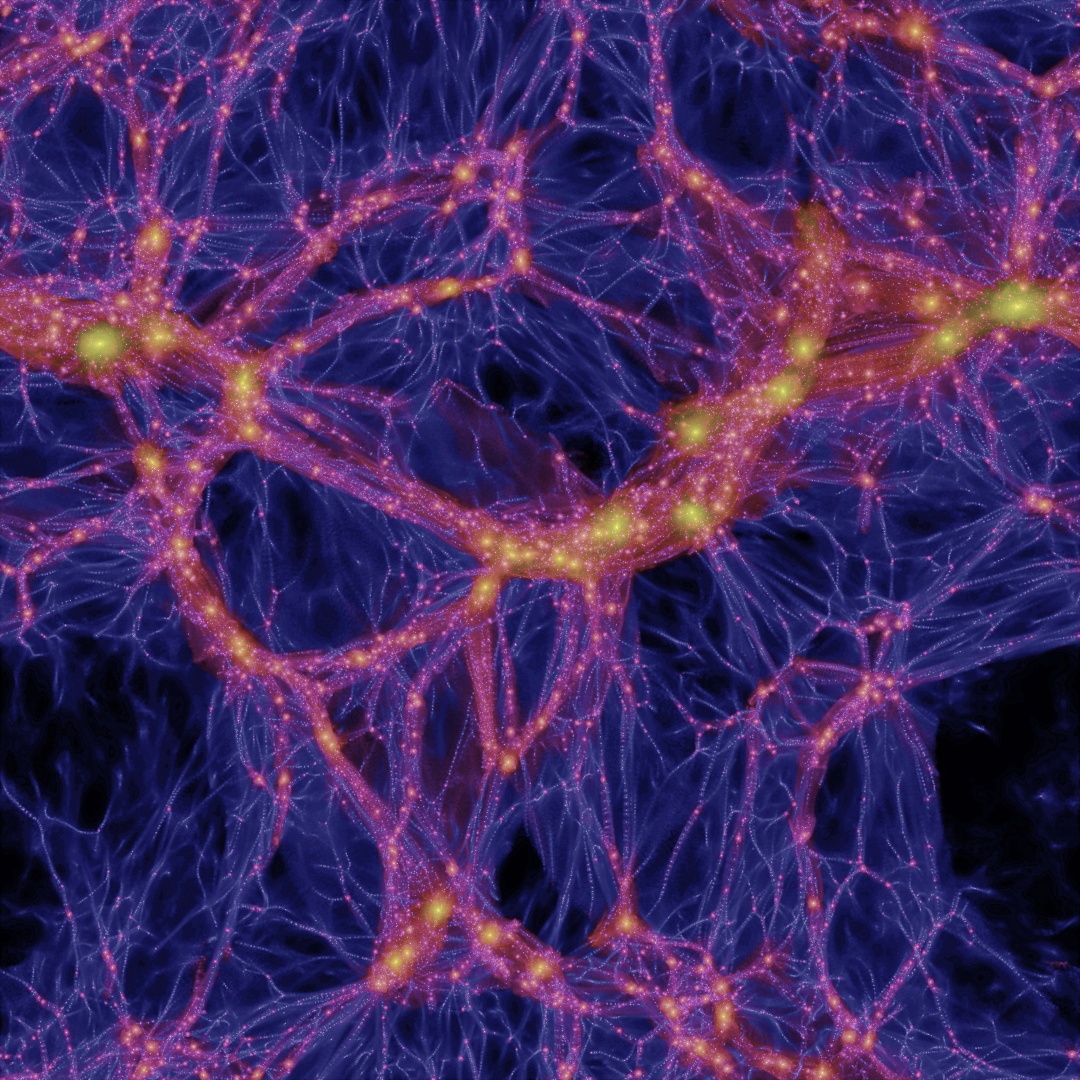}\llap{\makebox[\wd1][l]{\raisebox{-0.1\wd1}{\includegraphics[width=0.36\textwidth]{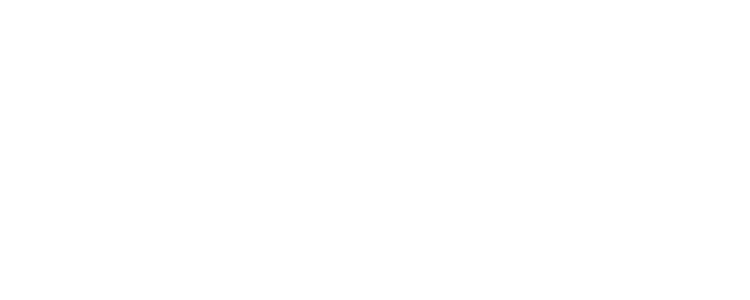}}}}
  \caption{Visualization of present-day DM distribution of the two simulations analysed here, CDM (left) and WDM (right). Each image is $L=34.4$ Mpc on a side. Image intensity indicates DM comoving density, and colour indicates velocity dispersion, with blue denoting low-velocity dispersion $(<5 \rm{km s^{-1}})$ and yellow high-velocity dispersion $(>200 \rm{km s^{-1}})$. The switch from a CDM to a WDM power spectrum clearly preserves large-scale structure but erases small-scale fluctuations.}
  \label{fig:sims}
\end{figure*}

\subsection{Numerical simulations}
\label{sec:sim}

We analyse a pair of high-resolution cosmological hydrodynamical simulations adopting initial conditions with matched phases but differing power spectra, one corresponding to CDM and one to WDM. The simulations were evolved with the EAGLE galaxy formation model \citep{Schaye2015, Crain2015}, which comprises a suite of subgrid models built into a modified version of the $N$-body Tree-PM smoothed particle hydrodynamics (SPH) code GADGET-3 \citep[last described by][]{Springel2005}. The simulations were evolved with mass resolution $8\times$ better than the flagship Ref-L100N1504 EAGLE simulation, yielding a baryon particle mass of $m_{\rm g} = 2.26\times 10^5\,{\rm M}_\odot$ and DM particle mass $m_{\rm dm} = 1.2\times 10^6\,{\rm M}_\odot$, and hence adopt the `Recal' model \citep[see][for further discussion]{Schaye2015}. The CDM simulation, introduced by \citet{Bastian2020} and recently examined in the study of \citet{2023arXiv231100041M}, is a $(34.4\,\mathrm{Mpc})^3$ volume realised at the same resolution as the Recal-L025N0752 simulation introduced by \citet[][thus corresponding to $N=1034^3$]{Schaye2015} and adopting the same cosmogony \citep[that of the][]{Planck2014}. This simulation also includes the E-MOSAICS globular cluster formation and evolution model \citep{2018MNRAS.475.4309P,2019MNRAS.486.3134K}, implemented as subgrid routines within EAGLE, but since these routines impart no `back-reaction' on the galaxy properties we do not discuss them here. 

The WDM counterpart simulation, which we introduce in this paper, adopts initial conditions generated using a modified power spectrum (see Sec.~\ref{sec:wdm}), but which are otherwise identical to those of the CDM simulation. The WDM simulation uses the same galaxy formation model adopted as the CDM simulation, with no recalibration of the feedback parameters: it nevertheless yields a present-day galaxy stellar mass function consistent with observations (Oman~et~al. in preparation). To the best of our knowledge, the WDM simulation is the first cosmological, hydrodynamical simulation of a reasonably representative cosmic volume within a WDM cosmogony that is broadly consistent with extant observational constraints. Both simulations adopt a Plummer-equivalent gravitational softening length of $\epsilon_{\rm com}=1.33~\rm{ckpc}$, limited to a maximum proper length of $\epsilon_{\rm prop}=0.35~\rm{ckpc}$.

A key aim of our paper is to compare the dwarf galaxy counts of the WDM simulation with those reported by \citet{Fattahi2020}. Those authors analysed the APOSTLE zoom simulations \citep{Fattahi2016,Sawala2016a} of LG analogues, considering galaxies with stellar mass $>10^{5}$~\Msun. The mass resolution of the simulations considered here is $\simeq 10\times$ poorer than the APOSTLE simulations at their resolution level `L1'. We must, therefore, be cautious when comparing galaxy counts at a mass scale close to the resolution limit of the simulations. \citet{Schaye2015} showed that resolution-related sampling effects result in an over-abundance of galaxies in EAGLE with fewer than 100 baryonic particles. \citet{Lovell2020c} found that the total stellar mass varies significantly with resolution: massive galaxies, with the stellar mass $M_{*} > 10^9$\Msun, form over twice as many stars at medium resolution than at high resolution, while the opposite is true for dwarf galaxies, with the high-resolution stellar mass in the range $M_{*} = [10^7,10^8]$~\Msun. We discuss convergence issues in the results and conclusions where relevant.

\subsection{WDM model}
\label{sec:wdm}

The WDM candidate sterile neutrino is characterized by three parameters: its mass, $M_{\rm s}$, the mixing angle $\theta_1$, and the lepton asymmetry, $L_6$, which we define as $L_6 \equiv 10^6 (n_{\nu_\rmn{e}} - n_{\bar\nu_\rmn{e}} )/s$, where $n_{\nu_\rmn{e}}$ is the lepton number density, $n_{\bar\nu_\rmn{e}}$ the anti-lepton number density and $s$ the entropy density. In principle, setting the value of two of these parameters uniquely determines the value of the third in order to obtain the correct DM abundance. Performing this calculation in practice is complicated by uncertainties in the computation of the lepton asymmetry at a fixed $M_\rmn{s}$ and $\theta_1$ \citep{Ghiglieri2015, Venumadhav2016, Lovell2023}. We adopt $\sin^2(2\theta_1)=2\times10^{-11}$ and $M_{\rm s} = 7.1~\rm{keV}$, as this model is the warmest model consistent with the decay interpretation of the 3.55~keV line. We compute the lepton asymmetry and the momentum distribution using the \citet{Lovell2016} implementation of \citet{Laine2008}, which gives the lepton asymmetry $L_6 = 11.2$. Note that more recent codes return very different $L_6$ values as well as different momentum distributions \citep{Ghiglieri2015, Venumadhav2016}. We then compute the linear matter power spectrum for this model using a modified version of the {\sc camb} Boltzmann solver code \citep{Lewis2000}.

WDM introduces a cutoff to the linear matter power spectrum, $P(k)$, at small scales, which significantly affects early structure formation. Computing the ratio of the two power spectra and taking its square root defines the transfer function: 
\begin{equation}\label{eq:transfer_func}
  T(k) \equiv \left(\dfrac{P_{\rm{WDM}}(k)}{P_{\rm{CDM}}(k)} \right)^{1/2}.
\end{equation}
The difference between CDM and WDM can then be parametrized using the half-mode mass, $M_{\rm{hm}}$, defined as:
\begin{equation}\label{eq:Mhm}
  M_{\rm{hm}} = \dfrac{4\upi}{3} \bar\rho \left(\dfrac{\upi}{k_{\rm{hm}}}\right)^{3} ,
\end{equation}
where $\bar\rho$ is the average density of the Universe, and $k_{\rm{hm}}$ is the half mode wavenumber, which is the scale where the transfer function drops by a factor of 2 and can therefore be thought of as the characteristic scale of the damping. We compute the half-mode mass for our model to be $M_{\rm{hm}}=6.3\times10^8$\Msun. This is equivalent to the mass of a thermal relic particle $m_{\rm{th}}=2.8 \rm{keV}$ using the approximation of \citet{Viel2005}.

Visualisations of the CDM and WDM present-day matter distributions in the two simulations are shown in Fig.~\ref{fig:sims}. The images demonstrate how WDM preserves the characteristic CDM matter distribution on large scales, but smooths density fluctuations on small scales. The projected density is encoded using brightness and the projected average three-dimensional velocity dispersion using the colour. The location of the massive haloes and the structure of the filaments that join them together are identical in the two models, highlighting that the formation time of massive haloes does not change significantly in response to the power spectrum alteration. The thermal motions of WDM particles at very early times erase low-mass structures leading to a paucity of low-mass haloes compared to CDM.

It has been shown that WDM simulations contain spurious haloes, which are numerical/resolution-dependent artefacts and, therefore, are not predictions of the underlying physical model. In simulations in which the initial power spectrum has a resolved cutoff, the small-scale structure is seeded in part by the discreteness of the particle set and generates spurious subhaloes. 
A significant fraction of spurious haloes can be identified and removed by performing a mass cut below a resolution-dependent scale, as suggested by \citet{wang2007}:
\begin{equation} \label{eq:Mlim}
  M_{\rm{lim}} = 10.1 \bar \rho d k_{\rm{peak}}^{-2},
\end{equation}
where $d$ is the mean interparticle separation and $k_{\rm{peak}}$ is the spatial frequency at which the dimensionless power spectrum, $\Delta(k)^2$, has its maximum. However, we need to avoid removing the genuine haloes below this mass scale. \citet{Lovell2014} demonstrated that one can further discriminate between genuine and spurious subhaloes by making a cut based on the shapes of the initial Lagrangian regions from which WDM haloes form. They found that, compared to genuine haloes, spurious candidates tend to have much more flattened configurations in their initial positions. We follow the methodology developed by \citet{Lovell2014} to identify spurious haloes and exclude them from the halo catalogues of the WDM simulation; we describe this method briefly below. 

The initial conditions specify the sphericity of haloes, which is defined as the axis ratio, $s=c/a$, of the minor to major axes in the diagonalized moment of inertia tensor of the initial particle load. The sphericity cut is made such that 99~per~cent of the CDM haloes containing more than 100 particles at the half-maximum mass snapshot lie above the threshold. We remove all haloes and subhaloes with the sphericity $s_{\rm{half-max}}<0.2$ and $M_{\rm{max}} < M_\rmn{cut} = 0.5 M_{\rm{lim}}$, where $M_{\rm{max}}$ is the maximum mass attained by a halo during its evolution. Using Eq.~\ref{eq:Mlim} we find $M_{\rm{cut}}$ for this simulation to be $M_{\rm{cut}}=1.6\times 10^8$~\Msun. This is chosen so as to identify a halo at a time well before it falls into a larger host, after which point its particles are subject to tidal stripping, and thus, some information about the initial conditions region may be lost. The factor of 0.5 is calibrated by matching between resolutions in the WDM Aquarius simulations \citep{Lovell2014}.  

\subsection{The LG galaxy sample}
\label{sec:sample}

Our primary source of LG galaxies is the most recent version of the LG catalogue of \citet{McConnachie2012}, with the addition of Crater~II \citep{Torrealba2016} and Antlia~II \citep{Torrealba2018} this catalogue gives 76 galaxies within 3~Mpc of the MW that have $M_{*}>10^{5}$~\Msun. We extended the catalogue with galaxies from the online Extragalactic Distance database\footnote{\url{edd.ifa.hawaii.edu}} \citep{Tully2009}. We use the {\em B}-band magnitude from the database and the stellar mass-to-light relations given in table~2 of \citet{Woo2008} to compute the stellar masses. This results in the addition of 12 dwarf galaxies to the LG, for a total of 88 dwarf galaxies. 

One complication of the observations is that some $\pm15$ degrees of the sky is obscured by the Galactic Plane in a region known as the `zone of avoidance'. We, therefore, implement the correction due to the zone of avoidance derived by \citet{Fattahi2020}. This correction assumes that the dwarf galaxy number density in the zone of avoidance is the same as the outside region; it increases the number of field dwarf galaxies by 8, to a total of 96 galaxies with $M_{*}>10^{5}$~\Msun.

\section{Results}
\label{sec:results}

\begin{figure*}
  \centering
  \includegraphics[width=1\textwidth, trim={0cm 0cm 0cm 0cm}]{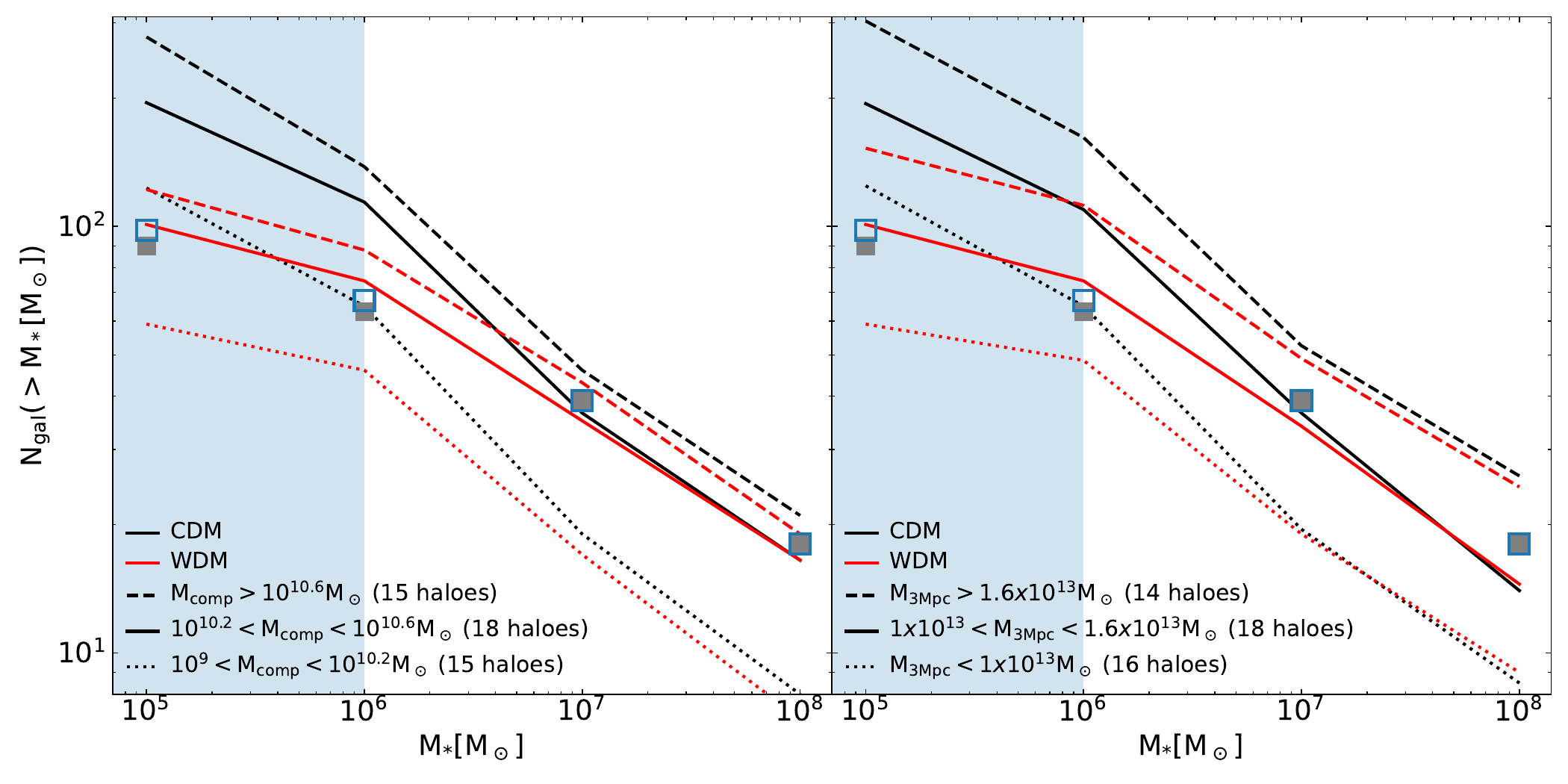}
  \caption{Cumulative stellar mass functions for MW-analogue haloes in different environments. The black and red lines correspond to the number of galaxies around MW-like haloes in CDM and WDM models, respectively. Grey solid squares represent the observed number of galaxies, while the open squares take into account the number of galaxies expected in the zone of avoidance. In the left-hand panel, different line styles indicate the highest mass companion of the main halo: $M_\rmn{comp}=[10^{9.0},10^{10.2}]$\Msun is shown as dotted lines, $M_\rmn{comp}=[10^{10.2},10^{10.6}]$\Msun are shown as solid lines and $M_\rmn{comp}>10^{10.6}$\Msun are presented as dashed lines. In the right-hand panel, different styles of lines indicate the total mass within 3~Mpc of the main halo. Dotted lines indicate $M_\rmn{3Mpc}<1\times10^{13}$~\Msun, solid lines represent the total mass in the range $M_\rmn{3Mpc}=[1,1.6]\times10^{13}$~\Msun and $M_\rmn{3Mpc}>1.6\times10^{13}$~\Msun is shown as dashed lines. The shaded regions indicate the galaxies in simulations with up to 10-star particles.}
  \label{fig:Ngal}
\end{figure*}

Our goal is to understand how the number of galaxies changes between the CDM and WDM models as a function of stellar mass and the local environment. As stated above, the cutoff in the WDM power spectrum causes not only a decrease in the number of low-mass haloes but also a delay in their formation time. The latter is related to where haloes are located, whether in cosmologically overdense or underdense regions \citep{Lovell2024}.

Our more specific interest is the impact of free-streaming on LG dwarfs and to understand how the DM content changes from the overdense regions immediately surrounding the MW and M31 to the underdense LG outskirts. This is complicated by two factors: the low statistics of galaxy counts associated with low-density regions of the LG and also that we do not have a large set of LG analogues with MW-M31 halo pairs at the measured MW-M31 separation. Therefore, we approach this problem in two stages. First, in Section~\ref{subsec:LGDwarfCounts}, we identify a series of MW-analogue haloes, broadly defined, and compute the galaxy counts in their vicinity. Second, we generate a series of randomly-positioned 3~Mpc-radius volumes with a variety of densities to identify how galaxy properties change from high-density regions to low-density regions, and present the results in Section~\ref{subsec:DwarfEnvironment}.

\subsection{LG dwarf galaxy counts in CDM and WDM}
\label{subsec:LGDwarfCounts}

We begin by building a catalogue of MW-mass haloes in the simulations. \citet{Callingham2019} estimate the mass of the MW halo to be $M^{\rm MW}_{200} = 1.17^{+0.21}_{-0.15}\times 10^{12}$~\Msun, and more broadly the mass is expected to be in the range $[0.8,2.4]\times10^{12}$~\Msun \citep{Watkins2019, Fritz2020, Karukes2020}, where the virial mass $M_{200}$ of the halo defined as the mass contained within a sphere of radius $r_{200}$ whose enclosed average density is 200 times the critical density. We are interested in systems that may contain a single MW-analogue halo or a MW-M31 pair within the parent FoF group, therefore we select FoF catalogue haloes that have $M_{200}$ in the range from the \citet{Callingham2019} $1\sigma$ lower limit to $2\times$ the \citet{Callingham2019}  $1\sigma$ upper limit, which is $M_{200}=[1.02,2.76]\times10^{12}$~\Msun. This criterion returns 48 systems. 
We draw spheres of radius 3~Mpc around these haloes and count the number of enclosed galaxies in the CDM and WDM simulations. 

We also follow \citet{Fattahi2020} and consider how the dwarf galaxy number density is affected by the presence of a massive companion galaxy like M31 in LG analogues, as well as by the overall mass of the LG. We therefore compute the total mass within each 3~Mpc volume, $M_\rmn{3Mpc}$, and identify each MW-analogue's most massive companion galaxy within 3~Mpc. We label the stellar mass of this companion $M_{\rm comp}$. We split the 48 haloes between 3 bins in $M_\rmn{3Mpc}$ and $M_\rmn{comp}$, with the bins chosen to have roughly equal numbers of MW analogues per bin. We compute the cumulative radial stellar mass function for each of the 48 volumes, where the stellar mass, $M_{*}$, is defined as the gravitationally bound mass of all star particles associated with each galaxy, and then calculate the median cumulative radial mass function for each of the three $M_\rmn{3Mpc}$ and the three $M_\rmn{comp}$ bins, then plot the results in Fig.~\ref{fig:Ngal}. The cumulative number of observed galaxies in our observational sample is indicated with solid squares, and the correction for the zone-of-avoidance is indicated with open squares. 

The simulations that we use have a star particle mass of $M_{*}=10^5$\Msun, which is equivalent to the smallest galaxies that we consider in this paper. We, therefore, indicate the stellar mass range of galaxies with fewer than 10-star particles in Fig.~\ref{fig:Ngal} as shaded regions to show where resolution introduces a particularly strong degree of uncertainty in our results. However, many gas particles are required to form a star particle, and therefore, the presence of a galaxy in the relevant subhaloes is well established; it is only the particular stellar mass that is uncertain. The utility of this plot is, therefore, in showing the general trend in the difference between CDM and WDM; we discuss the requirements for precise predictions later in this section.

As seen in the left panel of Fig.~\ref{fig:Ngal}, the WDM model haloes that best match the observed radial number density (of galaxies with $M_{*}>10^{7}$~\Msun) have $M_\rmn{comp}=[10^{10.2},10^{10.6}]$~\Msun and $M_\rmn{3Mpc}=[1-1.6]\times 10^{13}$~\Msun. The same is true for CDM, as in this regime, the number of predicted dwarfs is nearly identical for the two models. However, below $M_{*}=10^{7}$~\Msun, the number of galaxies in the CDM simulation increases almost linearly (in log-log space) while the WDM curve flattens markedly. Remarkably, the WDM turnover occurs at a very similar mass to that in which the observations diverge from CDM.  
For MW-analogues with a companion galaxy $M_\rmn{comp}=[10^{10.2},10^{10.6}]$\Msun, the number of galaxies in the CDM simulation is $92$~per~cent and $53$~per~cent higher than in the WDM simulation at $M_{*}=10^5$~\Msun and $M_{*}=10^6$~\Msun, respectively. 

The right-hand panel of Fig.~\ref{fig:Ngal} shows that for galaxies with total mass within 3~Mpc in the mass range $M_\rmn{3Mpc}=[1,1.6]\times10^{13}$\Msun, the CDM simulation yields $92$~per~cent and $46$~per~cent more galaxies than the WDM simulation at $M_{*}=10^5$~\Msun and at $M_{*}=10^6$~\Msun, respectively. That the greatest difference between CDM and WDM is found for low stellar mass systems follows naturally from the familiar WDM suppression of small-scale perturbations, especially at mass scales around and below the half-mode mass, $M_{\rm{hm}}$. 

We note that our CDM mass functions are somewhat steeper than those of \citet{Fattahi2020}. Our simulations differ from theirs in two relevant ways: (i) their volumes are tailored specifically to the MW-M31 system while ours are restricted to what is available in the box, and (ii) their simulations used the EAGLE Reference galaxy formation model while ours use the Recal model. The primary takeaway from our result is simply that the WDM predictions very roughly track where the observations where the CDM predictions instead diverge, despite the fact that the WDM model was not tuned to reproduce this result. A much more careful treatment of both MW-M31 pair selection and the sub-kpc galaxy formation physics will be required to make precise predictions that could constrain either model.

\begin{figure*}
  \centering
  \setkeys{Gin}{width=0.33\linewidth}
	\includegraphics{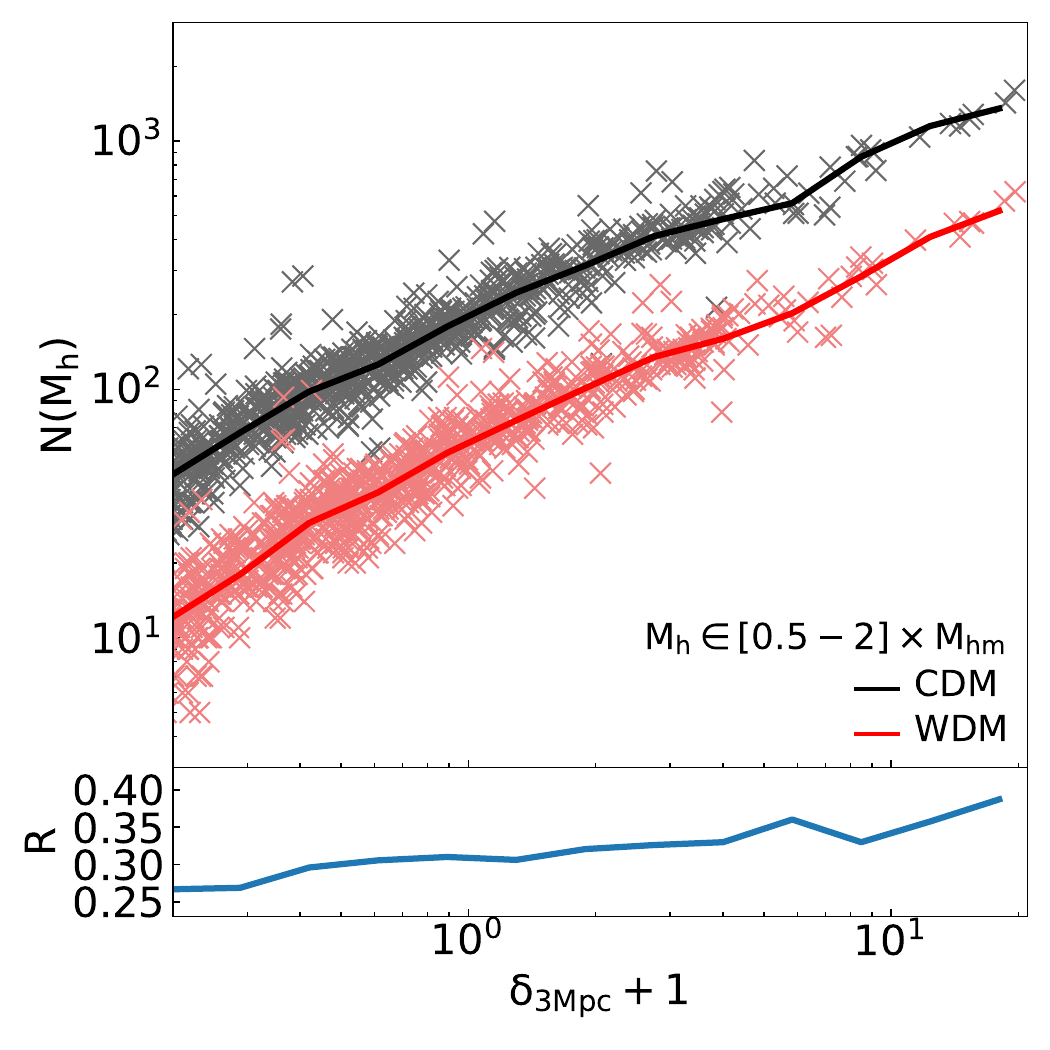}
  \hfill
	\includegraphics{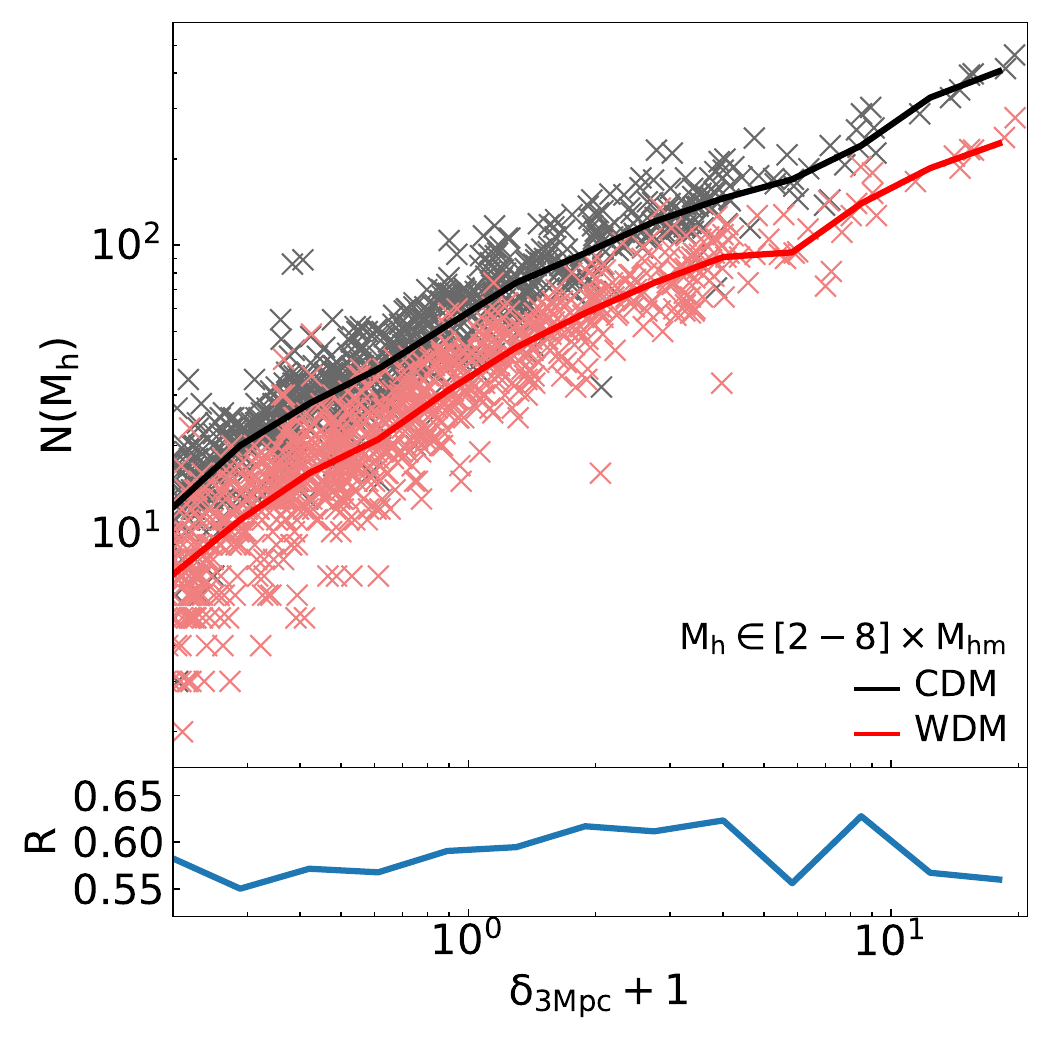}
  \hfill
	\includegraphics{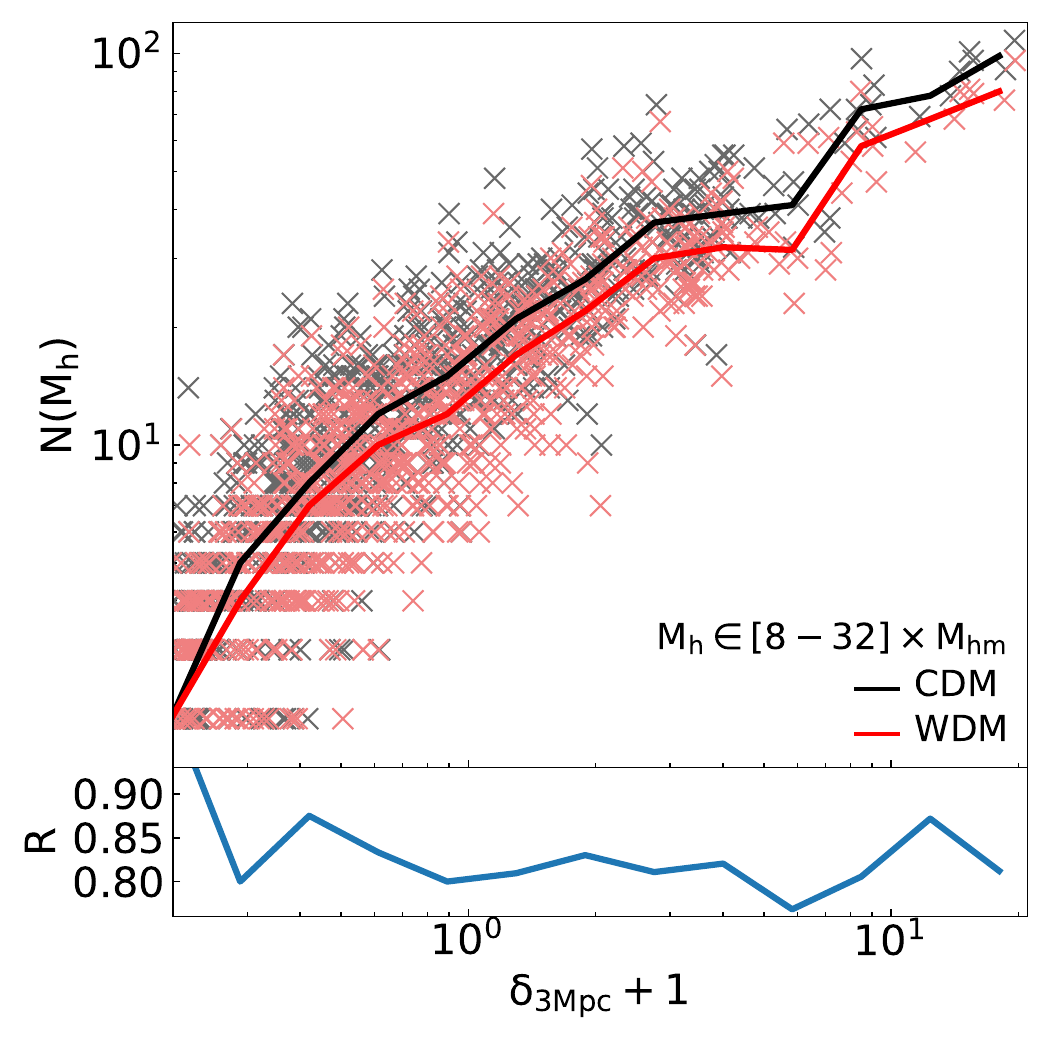}
  \caption{Number of haloes as a function of overdensity.
  The left, middle and right-hand panels correspond to $[0.5,2]\times M_\rmn{hm}$, $[2,8]\times M_\rmn{hm}$, and $[8,32]\times M_\rmn{hm}$ halo mass ranges, which corresponds to halo mass bins of $[3.15\times10^8 - 1.26\times10^9]$\Msun, $[1.26\times10^9 - 5.04\times10^9]$\Msun and $[5.04\times10^9 - 2.016\times10^{10}]$\Msun, respectively. Black and red colours indicate the CDM and the WDM counterparts, respectively. Crosses indicate the number of haloes in the randomly chosen 1000 volumes, and solid lines are the median relations. Each bottom panel shows the ratio of the WDM and CDM median relations. The left panel has the biggest slope of the ratio, while the middle has a minor increase, and the right panel is effectively flat. Note the vastly different y-axis ranges of the sub-panels.}

  \label{fig:NMhalo}
\end{figure*}

\subsection{Dwarfs and their environment}
\label{subsec:DwarfEnvironment}

Having shown that the number of dwarf galaxies differs significantly between CDM and WDM simulations for a fixed galaxy formation model and that the divergence grows at a mass scale that is highly relevant for comparisons to observations, we next turn to an exploration of how the number of dwarf galaxies arises from the combination of the dwarf halo abundance, the impact of reionization in preventing galaxy formation, and the stellar mass--halo mass relation. We consider all of these factors as a function of the matter density within 3~Mpc. 

We begin by analysing the impact of the different halo mass functions between the two models, specifically the change in the number of haloes as a function of the overdensity parameter, $\delta_\rmn{3Mpc}+1 \equiv \rho_\rmn{3Mpc}/\rho_\rmn{crit}$. We choose three dwarf halo mass bins: $[0.5,2] \times M_\rmn{hm}$, $[2,8]\times M_\rmn{hm}$, and $[8,32]\times M_\rmn{hm}$, which corresponds to halo mass bins of $[3.15\times10^8 - 1.26\times10^9]$\Msun, $[1.26\times10^9 - 5.04\times10^9]$\Msun and $[5.04\times10^9 - 2.016\times10^{10}]$\Msun, respectively. For our definition of halo mass, we use the gravitationally bound mass including all mass species, otherwise known as the dynamical mass, $M_\rmn{dyn}$. We use $M_\rmn{dyn}$ instead of $M_{200}$ here as our dwarf halo sample includes both isolated haloes and satellites. We count the number of haloes in 1000 randomly centred 3~Mpc-radius spheres in the WDM and CDM counterpart simulations. We include both isolated haloes and subhaloes, and make no distinction between them; for the remainder of this section, we refer to both types of objects as `haloes' for brevity. 

As shown in the left panel of Fig.~\ref{fig:NMhalo}, the number density of galaxies with mass around and below the half-mode mass is more strongly suppressed in the most underdense regions. The ratio $n_{\rm WDM}/n_{\rm CDM}$ is $\simeq 0.27$ at $\delta_\rmn{3Mpc}+1=0.2$ but increases to $\simeq 0.34$ at $\delta_\rmn{3Mpc}+1=10$. In the middle and the high mass ranges, shown in the centre and right-hand panels, the number of haloes in the two cosmogonies is similar, particularly so for the most massive haloes, where the WDM-to-CDM number density ratio is between $\simeq 0.80$ and $\simeq 0.90$. The difference between WDM and CDM, therefore, widens with decreasing local density, especially for halo masses around (or less than) the half-mode mass.
\begin{figure*}
  \centering
  \setkeys{Gin}{width=0.33\linewidth}
	\includegraphics{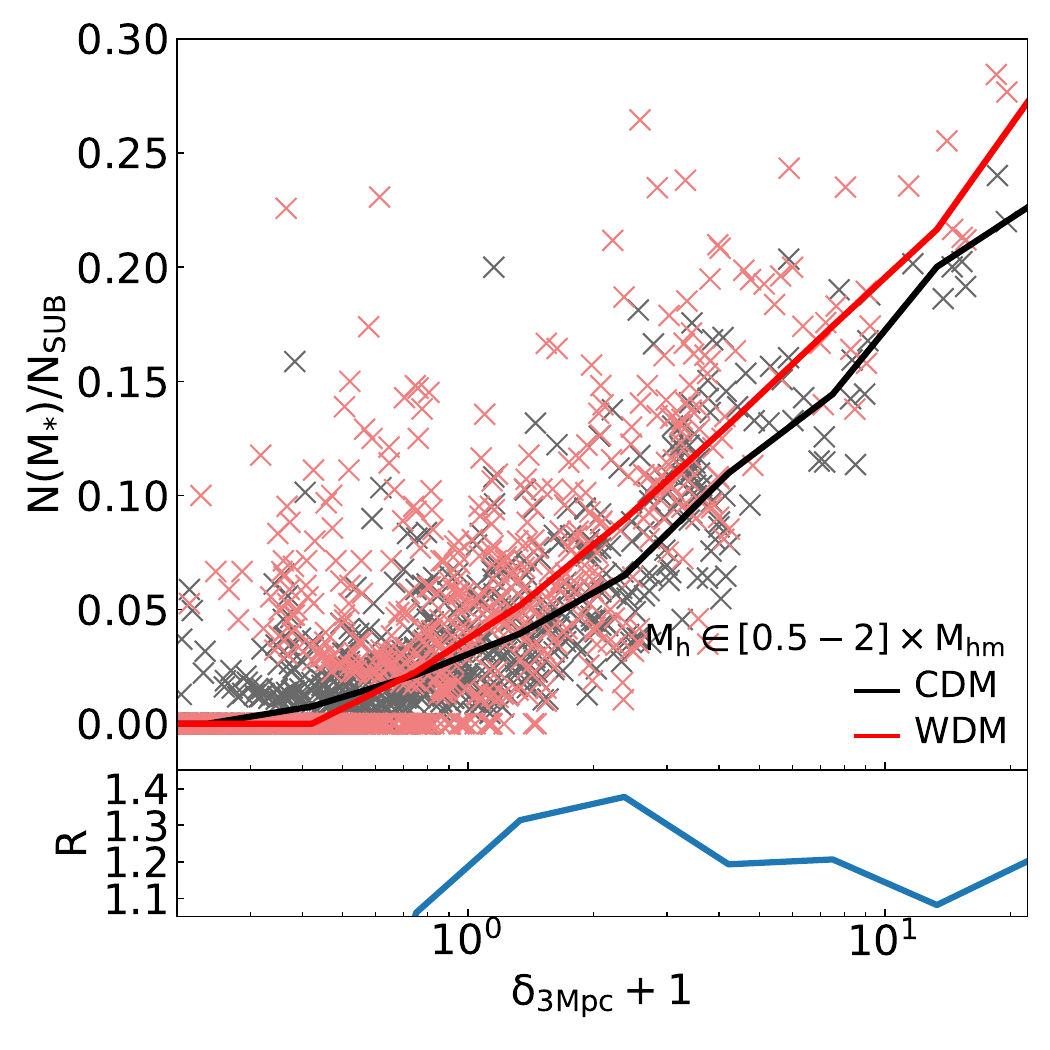}
  \hfill
      \includegraphics{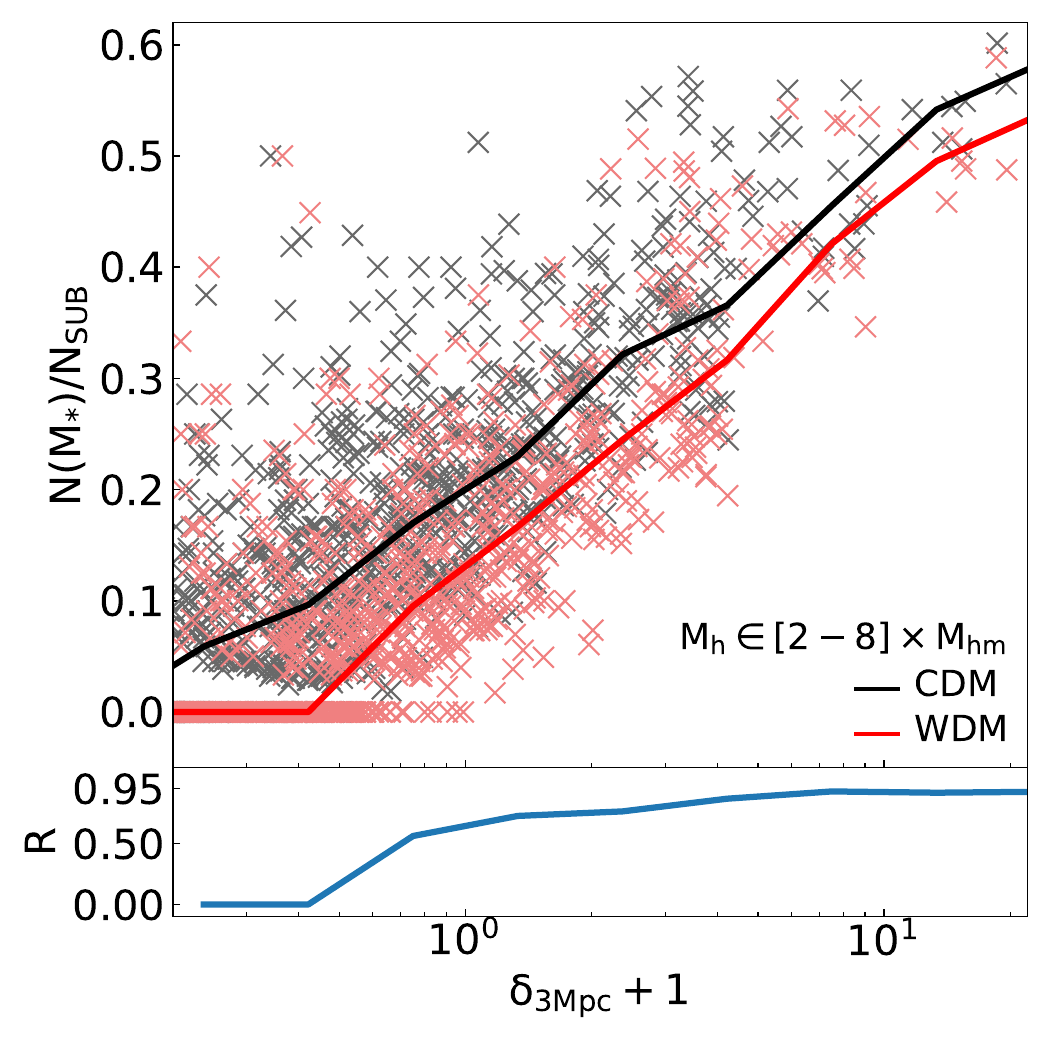}
  \hfill
	\includegraphics{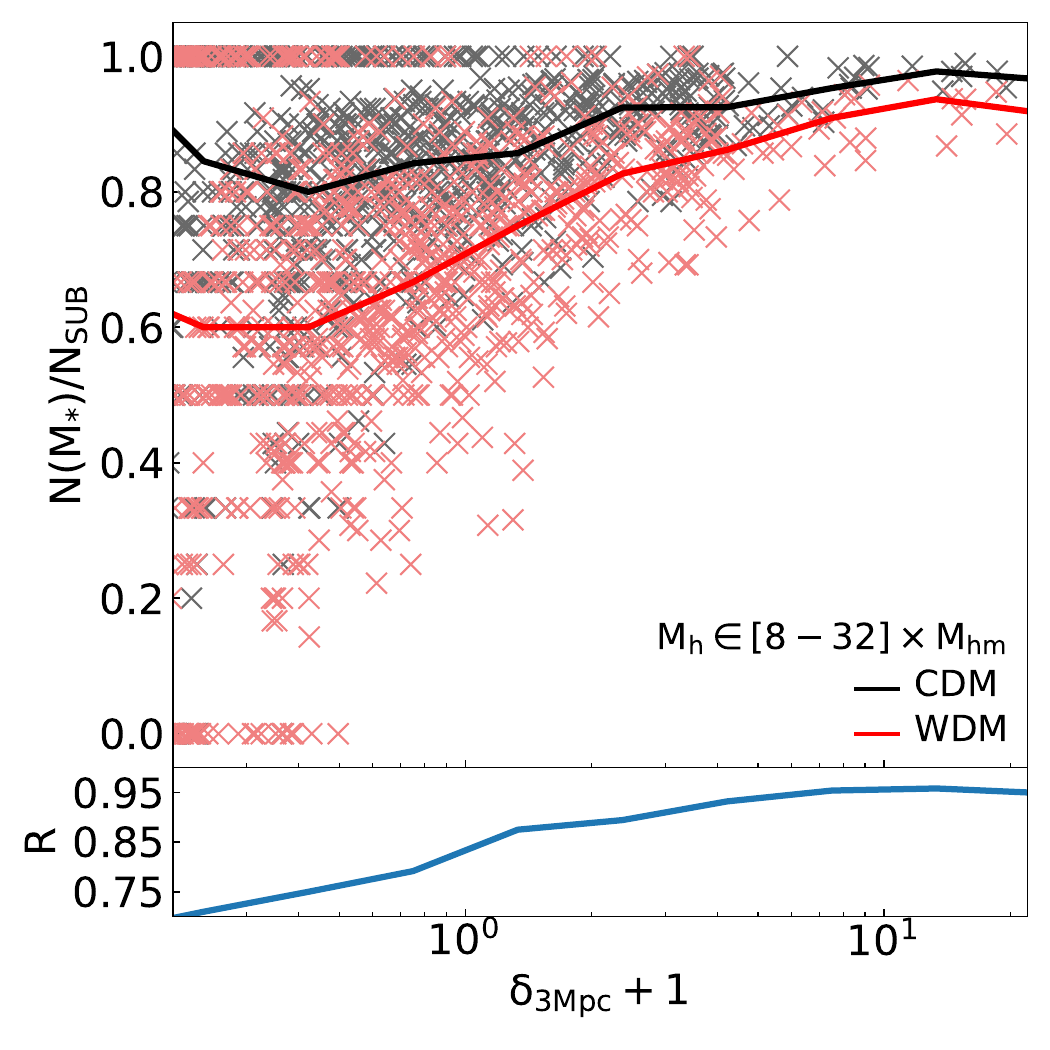}
  \caption{Luminous fraction as a function of overdensity. The left middle and right-hand panels correspond to $[0.5,2]\times M_\rmn{hm}$, $[2,8]\times M_\rmn{hm}$, and $[8,32]\times M_\rmn{hm}$ halo mass ranges, which corresponds to halo mass bins of $[3.15\times10^8 - 1.26\times10^9]$\Msun, $[1.26\times10^9 - 5.04\times10^9]$\Msun and $[5.04\times10^9 - 2.016\times10^{10}]$\Msun, respectively. Crosses indicate the luminous fraction in the randomly chosen 1000 volumes, and solid lines correspond to the median relations as a function of overdensity $\delta_{\rm{3Mpc}}+1$. Each bottom panel shows the ratio of WDM and CDM median relations.}

  \label{fig:LF}
\end{figure*}
We next consider the variation in the luminous halo fraction, which we define as the ratio of the number of haloes hosting at least one star particle to the total number of haloes within a given mass bin. We follow the template of Fig.~\ref{fig:NMhalo} and show the results in Fig.~\ref{fig:LF} as a function of $\delta_\rmn{3Mpc}+1$ for low-, intermediate- and high-mass haloes in the left, centre and right-hand panels, respectively. In CDM, the luminous fraction at the highest overdensities sampled by the simulations is $0.2$, $0.55$ and $1$ in the low, middle and high-mass ranges, respectively. These fractions then decrease monotonically from overdense to underdense regions. The ratio of the WDM-to-CDM relations in the middle and upper halo masses ranges becomes smaller towards less dense regions to the ratio of $0.5$ 
at $\delta_\rmn{3Mpc}+1=0.5$ for the middle mass range and $0.75$ at the same underdensity for the higher mass range. This behaviour results from the interplay of the delay in WDM halo formation and the impact of reionization: the delay in formation shifts the halo collapse time past the end of the reionization epoch, inhibiting and preventing the cooling of gas and the formation of a galaxy. The picture in the lowest halo mass bin is somewhat different, with a higher luminous fraction in WDM than CDM at all overdensities. This result can be explained by the width of the halo mass bin and the shape of the halo mass function. The luminous fraction of both CDM and WDM haloes at $2\times M_\rmn{hm}$ is higher than that at $0.5\times M_\rmn{hm}$, but the abundance of CDM haloes at the lower end of the bracket is much larger than that of either WDM haloes at that mass or of either model at $2\times M_\rmn{hm}$.

\begin{figure*}
  \centering
  \setkeys{Gin}{width=0.33\linewidth}
	\includegraphics{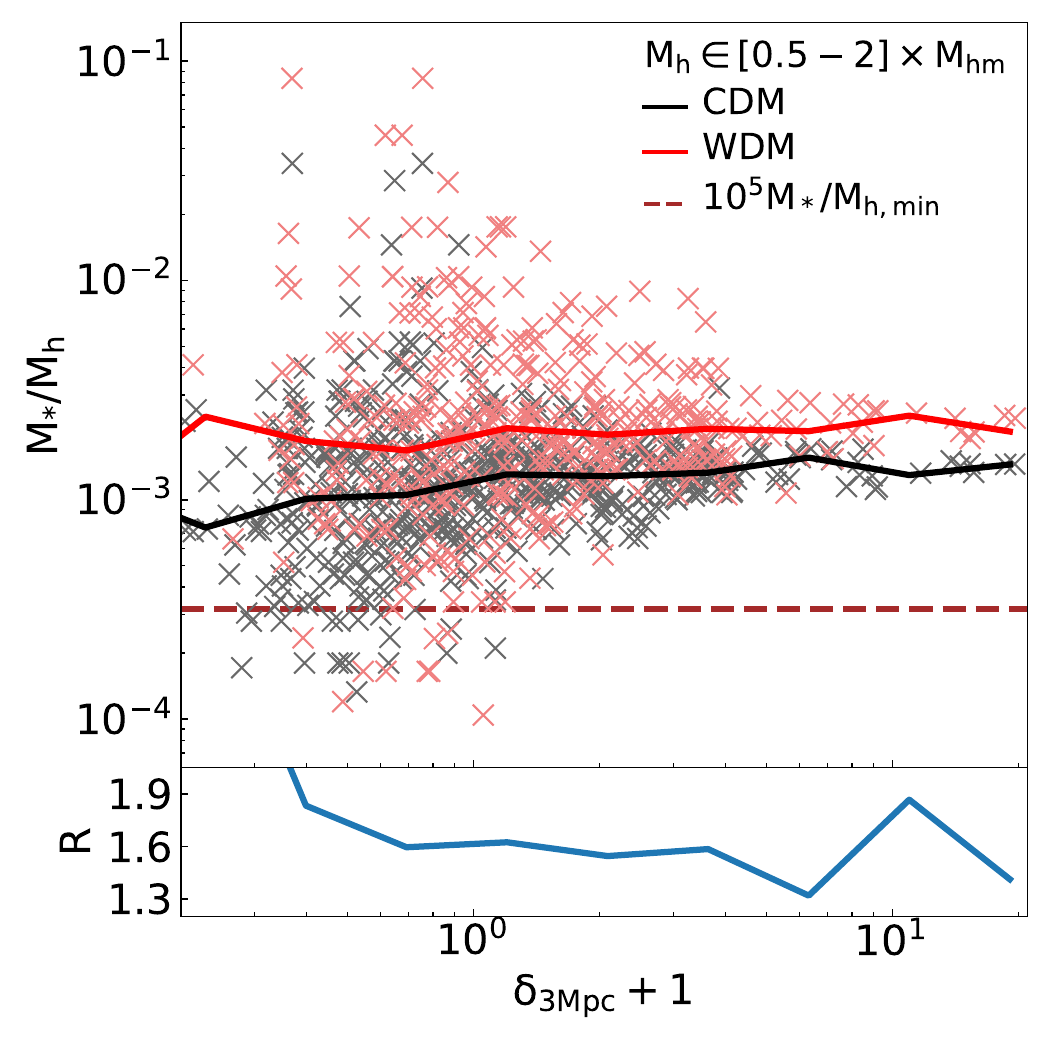}
  \hfill
	\includegraphics{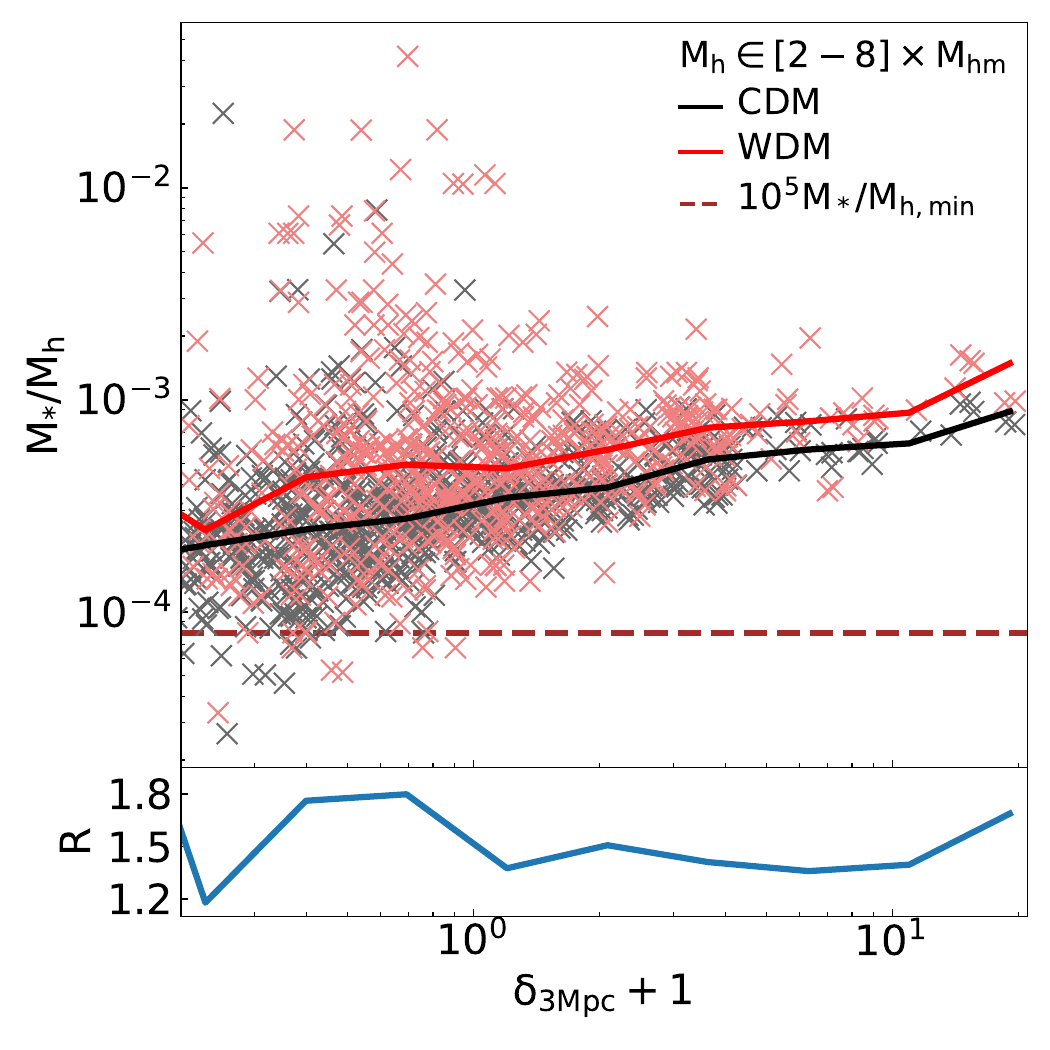}
  \hfill
	\includegraphics{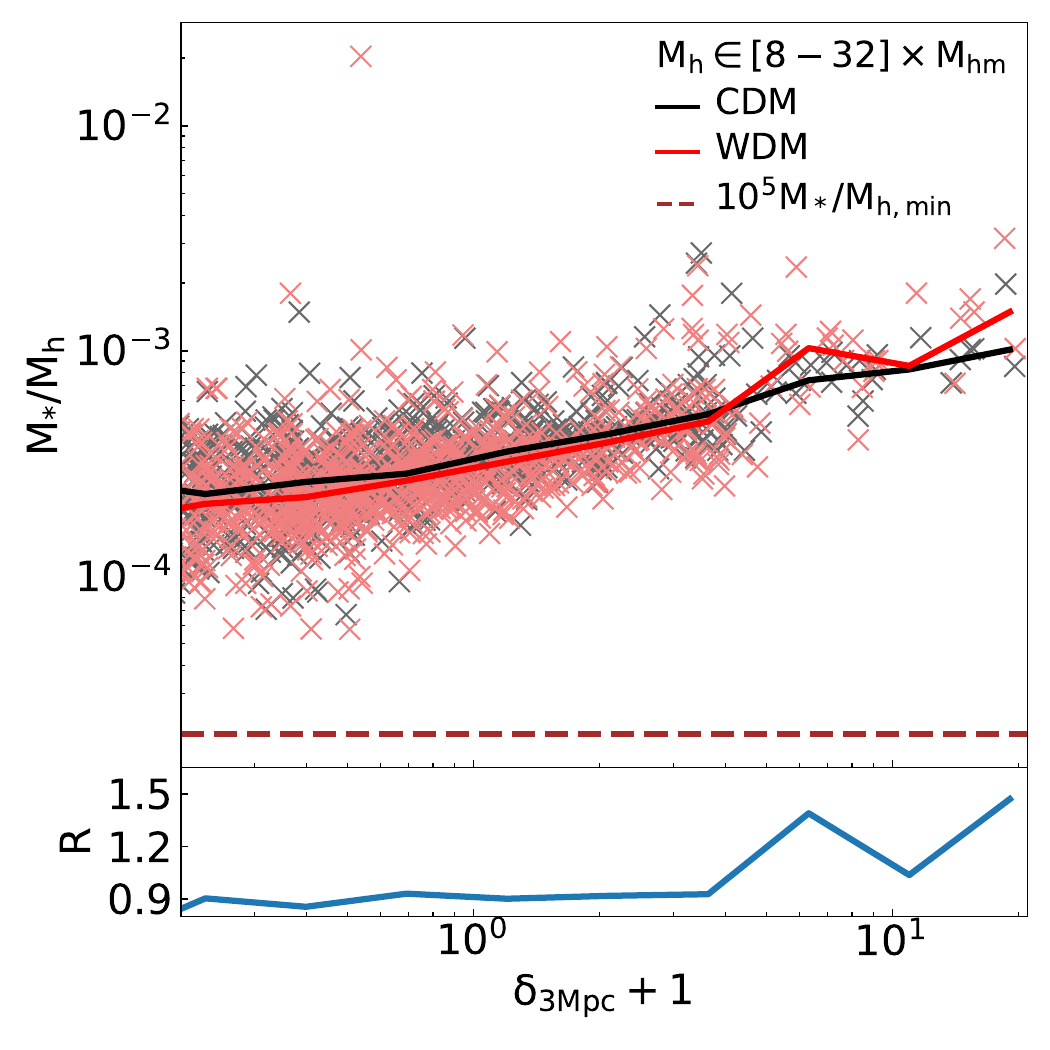}
  \caption{The stellar mass-halo mass relation as a function of overdensity. The left, middle and right-hand panels correspond to $[0.5,2]\times M_\rmn{hm}$, $[2,8]\times M_\rmn{hm}$, and $[8,32]\times M_\rmn{hm}$ halo mass ranges, which corresponds to halo mass bins of $[3.15\times10^8 - 1.26\times10^9]$\Msun, $[1.26\times10^9 - 5.04\times10^9]$\Msun and $[5.04\times10^9 - 2.016\times10^{10}]$\Msun, respectively. Crosses indicate the median values of the stellar mass-halo mass relation in the randomly chosen 1000 volumes, and solid lines correspond to the median relations. The horizontal dashed lines indicate the mass ratio associated with a single particle. Each bottom panel shows the ratio of the WDM and CDM median relations.}

  \label{fig:MsMh}
\end{figure*}

We have shown that the transition to lower-density regions decreases both the relative total number of haloes and the relative fraction of luminous haloes between WDM and CDM. The final property to consider is the stellar mass--halo mass relation. For those haloes that do form a galaxy, how many stars do they form? The monolithic collapse of WDM haloes \citep{Lovell2019} suggests that they may lose less of their gas to stellar feedback at very early times than is the case in CDM and thus yield a higher stellar mass--to--halo mass ratio. We, therefore, compute this ratio for all haloes with at least one star particle, and in Fig.~\ref{fig:MsMh} show the median of this ratio as a function of overdensity and halo mass, as per previous figures. 

The median stellar-to-halo mass ratios in WDM are consistently greater than those in CDM in the low and intermediate-mass bins, while the two are similar in the high-mass bin. Remarkably, the median ratio of the WDM to CDM stellar mass--halo mass relation increases from overdense to underdense regions. In the low-mass bin, the ratio of WDM to CDM median stellar mass per unit halo mass increases from 1.4 to 1.8 between $\delta_{\rm{3Mpc}}+1=5$ and $\delta_{\rm{3Mpc}}+1=0.4$; in the intermediate-mass bin the ratio increases from 1.4 to 1.8 for $\delta_{\rm{3Mpc}}+1=10$ to $\delta_{\rm{3Mpc}}+1=0.7$. We caution that many haloes in the low mass case only host a single particle and that the value of $M_{*}$ for any individual galaxy is highly unreliable for the reasons discussed in Section~\ref{sec:methods}. However, a similar result has been inferred from the same data set using mock {\sc HI} surveys (Oman~et~al. in preparation) and has also been shown for a power spectrum cutoff at high redshift in an independent galaxy formation model \citep{Lovell2019}. This result supports the hypothesis that star formation in WDM haloes in underdense regions can quickly catch up and surpass the CDM through a strong late-time ($z\sim8$) starburst \citep{Bose2016b}, which leads to a population of dwarf galaxies that is brighter than would otherwise be the case, including in underdense regions.

\begin{figure*}
  \centering
  \setkeys{Gin}{width=0.33\linewidth}
	 \includegraphics{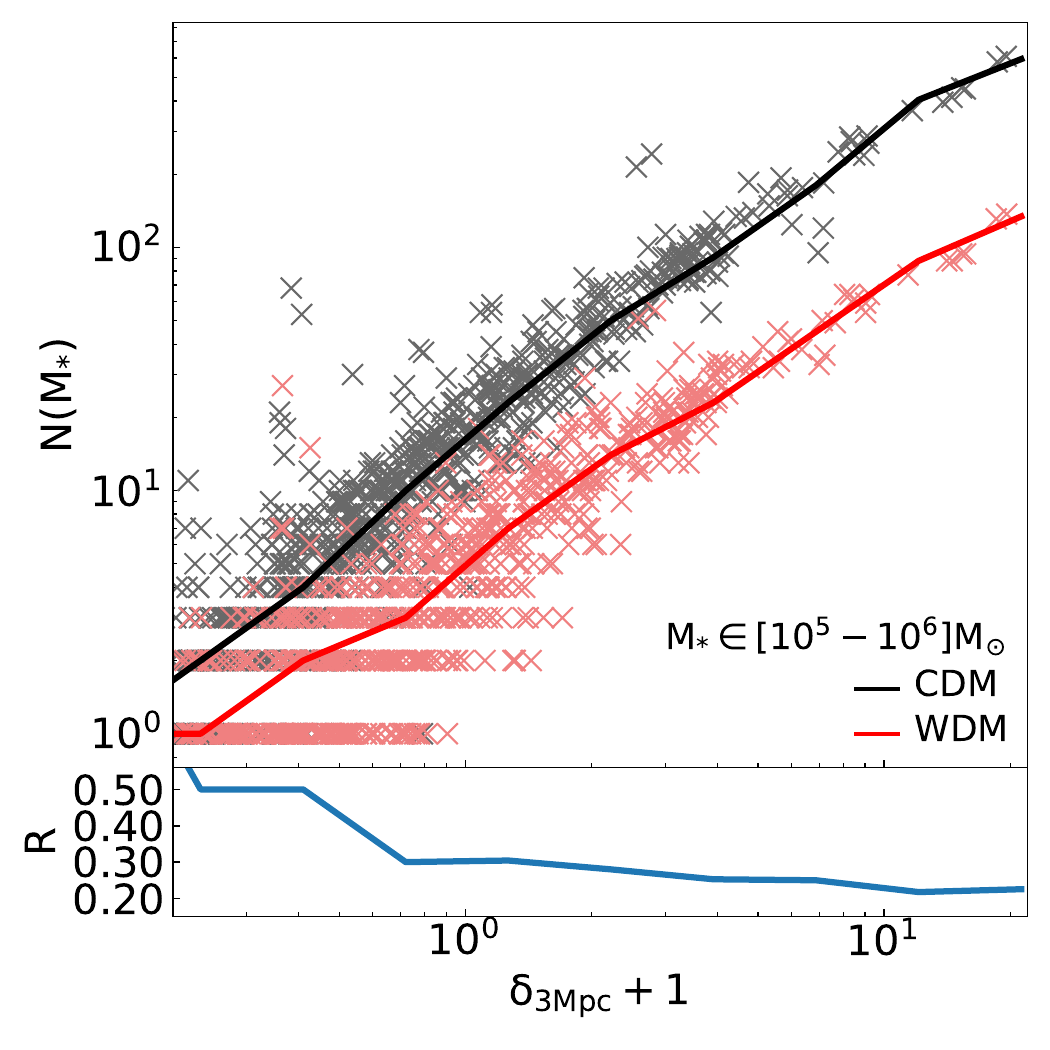}
 \hfill
	\includegraphics{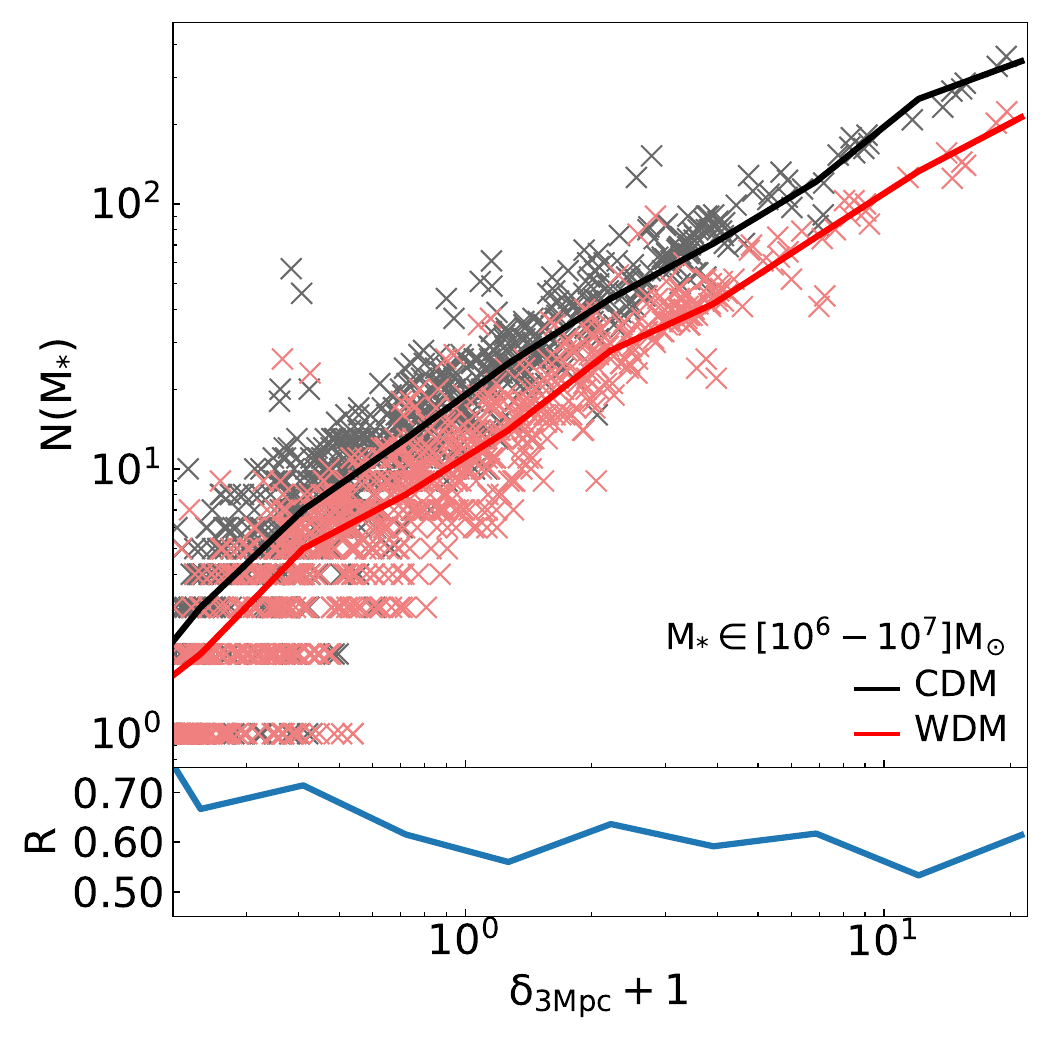}
  \hfill
	\includegraphics{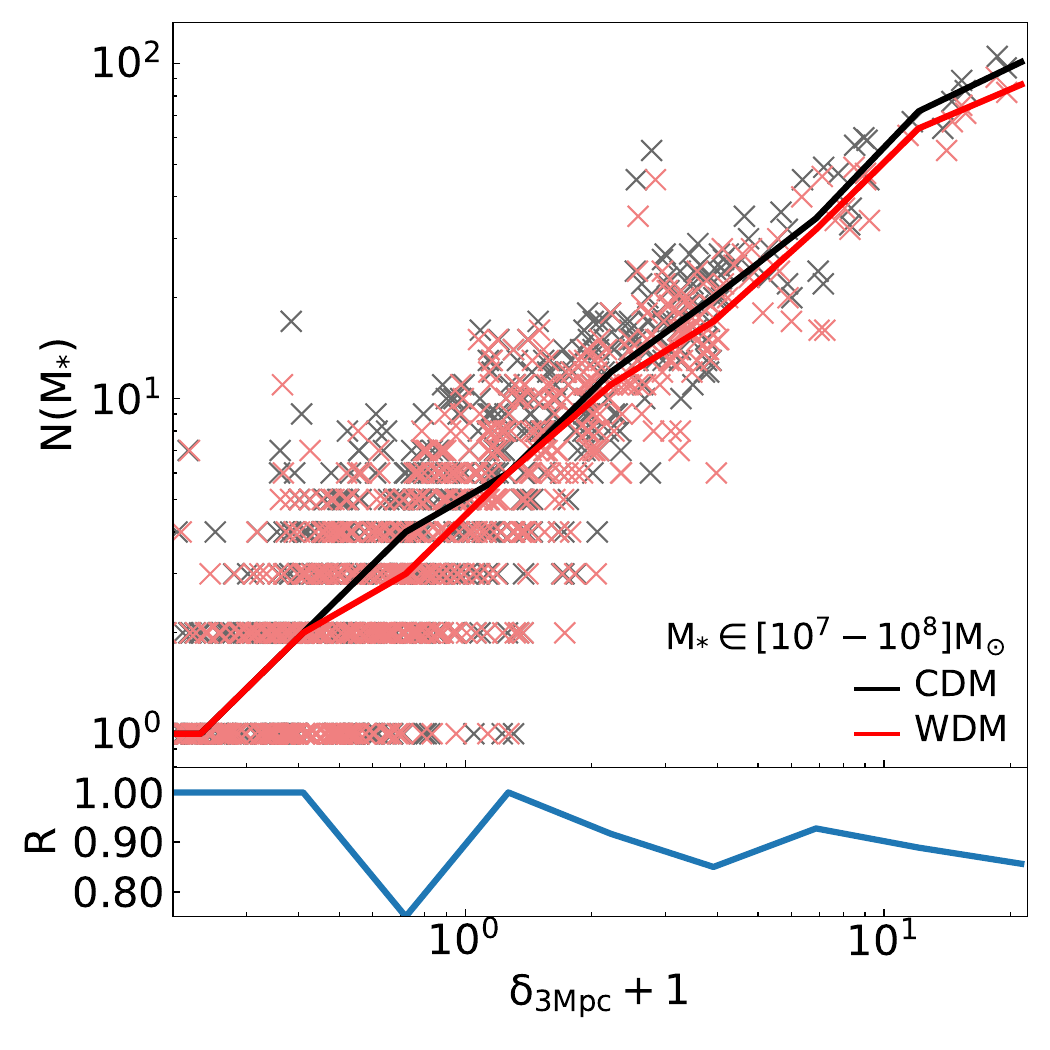}
  \caption{Number of galaxies as a function of overdensity. The left middle and right-hand panels correspond to $[10^5 - 10^6]$\Msun, $[10^6 - 10^7]$\Msun, and $[10^7 - 10^8]$\Msun stellar mass ranges, respectively.  Crosses indicate the number of galaxies in the corresponding 3Mpc radius volume. Solid red and black lines correspond to the medians in WDM and CDM counterparts according to the overdensity, $\delta_{\rm{3Mpc}}+1$ values. The bottom panels show the ratio of the WDM median to the CDM median values.}

  \label{fig:NMstar}
\end{figure*}

We have the three components that combine to provide the stellar mass function: the halo mass function, the luminous fraction, and the stellar mass--halo mass relation. We now analyse how the stellar mass function itself changes from high-density regions to low-density regions. We therefore consider the number density of luminous galaxies as a function of overdensity, now split into three stellar mass bins rather than three halo mass bins: $\left[10^{5}-10^{6}\right]$, $\left[10^{6}-10^{7}\right]$, and $\left[10^{7}-10^{8}\right]$~\Msun. We present the results in Fig.~\ref{fig:NMstar}.

The decreasing ratio of the WDM to CDM galaxy abundance 
at the highest sampled overdensities is more pronounced for low-mass galaxies, being $0.85$ 
in the highest mass bin and $0.60$ 
at $[10^{6},10^{7}]$~\Msun and $0.30$ 
at $[10^{5},10^{6}]$~\Msun. Of greater interest is the fact that this suppression differs from the halo abundance by remaining constant with decreasing overdensity rather than increasing, down to the regime where the resolution has a significant impact. Part of this result can likely be explained by the high luminous fraction of haloes with masses $8M_\rmn{hm}$ across all environments. However, even this class of haloes exhibits a large luminous fraction difference between CDM and WDM in underdense regions. Instead, we have a picture where the increase in the WDM stellar mass-halo mass relation with decreasing local density balances the impact of greater halo number density suppression and lower luminous fractions. The details of this process will be strongly dependent on the details of dwarf galaxy astrophysics at high redshift \citep{Shen2023}, including how reionization proceeds on small scales and how gas cools into low-mass haloes. We have shown that free-streaming can, in principle, lead to changes in galaxy counts with the local density. However, the degree of the change will depend strongly on astrophysical processes, and the magnitudes of their effects are poorly understood.

\section{Conclusions}
\label{sec:conclusions}

Experimental detection and consistency with galaxy observations are needed to claim that we have an authoritative description of DM. The CDM model has yet to be verified through any experimental detection of its DM candidate, and tensions remain with dwarf galaxy observations, including the abundance of dwarf galaxies in the LG. The WDM model may offer a remedy for both elements of this problem in that its sterile neutrino particle physics candidate has been potentially detected in X-ray emission \citep{Boyarsky2014,Bulbul2014,Hofmann2019}, and this detection sets a characteristic scale that can impact dwarf galaxy properties in a manner that becomes more pronounced in underdense regions.

In this paper, we examine how the switch from CDM to WDM changes the properties of dwarfs in underdense regions for a fixed galaxy formation model (one calibrated using CDM simulations). We select a WDM candidate sterile neutrino with a mass $M_{\rm s} = 7.1$~$\rm{keV}$, mixing angle $\sin^2(2\theta_1)=2\times10^{-11}$, and the lepton asymmetry $L_6 = 11.2$ as computed using \citet{Laine2008} and \citet{Lovell2016}. We identify MW-mass haloes broadly defined -- $M_\rmn{200}=[1-2]\times\rm{M}^{\rm{MW}}$ -- which yields 48 haloes with the virial mass in the range $M_{200} \in [1.02 - 2.76] \times 10^{12}$\Msun in CDM.

We measured the abundance of galaxies within 3~Mpc of each MW-analogue halo (Fig.~\ref{fig:Ngal}) and found that the cumulative stellar mass function is nearly identical between WDM and CDM at $M_{*}>10^7$\Msun. However, below this mass, the cumulative mass function becomes shallower for WDM, where this model predicts a smaller number of dwarf galaxies than CDM, at 50~per~cent of the CDM value for $M_{*}>10^{5}$~\Msun. Curiously, the turnover in WDM occurs at a similar mass scale to where the observations depart from the CDM prediction. Our results suggest that in the WDM model, the total mass within 3~Mpc is in the range $[1-1.6]\times 10^{13}~$\Msun, and the MW's most massive companion halo has a stellar mass higher than $>10^{10.2}$~\Msun. For the latter, we consider the  $[10^{10.2} - 10^{10.6}]$~\Msun and  $>10^{10.6}$~\Msun mass bins together since they are within the region-to-region scatter. Furthermore, the EAGLE model predicts a $\rm{M_* / M_h}$ relation for these mass ranges lower than is expected from observations \citep{Schaye2015}, thus an improved model may revise the values of $\rm{M_{comp}}$ upward. This result is within an order of magnitude of the inferred stellar mass of M31 estimated using optical and near-infrared imaging data \citep{Tamm2012}, and considering the correction, it can potentially return a better match to the observations.

One of the uncertainties in measuring the full population of LG dwarfs is depth completeness: could there be distant galaxies in the lower-local-density parts of the LG that we have not yet detected? The evolution of dwarf haloes between CDM and WDM is a function of local density: therefore, we study how the properties of dwarfs change in a series of 1000 randomly centred spheres or radius 3~Mpc. The ratio of halo number in WDM relative to CDM decreases significantly from high-density regions to low-density regions (Fig.~\ref{fig:NMhalo}), from $0.40$ 
at $\delta_\rmn{3Mpc}+1=20$ to $0.27$ 
at $\delta_\rmn{3Mpc}+1=0.2$ for haloes in the region of the half-mode mass.

The luminous fraction in the two models also diverges from the top of the density range to the bottom for $M_\rmn{dyn}>2M_\rmn{hm}$, as the increased collapse delay at small densities pushes the collapse to after the reionization threshold (Fig.~\ref{fig:LF}). However, we find that the stellar mass--halo mass relation of WDM haloes relative to CDM {\it increases} towards lower density regions for haloes of mass $<8M_\rmn{hm}$ (Fig.~\ref{fig:MsMh}), which we posit is due to the absence of energy injection from stellar feedback at $z>9$ as is the case in CDM. 
The net result is that the relative number of galaxies in the two models does not change between overdense and underdense regions: while both the relative number of haloes and the relative fraction of luminous fraction decrease towards lower densities, the concurrent increase in the stellar mass-halo mass relation compensates for the prior effects. 

We have demonstrated in this paper that the 7.1~keV sterile neutrino DM candidate has the potential to explain the purported deficit of faint LG dwarfs and that the population of these dwarfs should be the same between the two models independent of the local density. While we were able to take advantage of a pair of high-resolution simulations sufficiently large to yield a reasonably representative galaxy population, we did not select specifically for LG-analogue volumes in that they did not feature a pair of MW and M31-analogue haloes at their precise separation; therefore future work will need to focus on APOSTLE-style volumes to check that low-density regions in the LG do exhibit the expected behaviour in WDM. Moreover, our results were obtained for a specific sterile neutrino model and a specific galaxy formation model, and it will be crucial to relax these conditions in future work. First, uncertainties in both the measured value of $\sin^{2}(2\theta)$ and in the particle physics calculations lead to a factor of 6 uncertainty in the 3.55~keV line-compliant $M_\rmn{hm}$ value. Second, the EAGLE galaxy formation model was calibrated to reproduce the key properties of the present-day galaxy population and hence does not focus specifically on the particular challenge of dwarf galaxy formation during the epoch of reionization. The latter problem is beginning to be addressed with new simulations that follow early dwarf formation with radiative transfer \citep[e.g.][]{Shen2023}, while the former will require a dedicated particle physics phenomenology effort. Should the {\it XRISM} mission detect a series of X-ray detection of the 3.55~keV line with the properties predicted for DM decay, this will provide compelling extra motivation to pursue this kind of particle physics work in addition to modelling the faintest galaxies in the Universe in the WDM cosmology. 

\section*{Acknowledgements}
We thank Jes\'us Zavala for the helpful discussions and comments on the draft. TM and MRL acknowledge support by a Project Grant from the Icelandic Research Fund (grant number 206930). RAC was supported by a Royal Society University Research Fellowship during the initial development of this study. JP is supported by the Australian government through the Australian Research Council's Discovery Projects funding scheme (DP220101863).

\section*{Data Availability}

Researchers interested in access to the simulations used in this paper should contact R.~A.~Crain at R.A.Crain@ljmu.ac.uk.



\bibliographystyle{mnras}
\bibliography{main} 





\bsp	
\label{lastpage}
\end{document}